# Ultrafast carbon nanotube photodetectors with bandwidth over 60 GHz

Weifeng Wu[1], Fan Yang[2], Xiansong Fang[2], Xiang Cai[1,2], Xiaohui Liu[1], Fan Zhang[2,3,4]*, Sheng Wang[1,2]*

[1] Key Laboratory for the Physics and Chemistry of Nanodevices and Center for Carbon-Based Electronics, Department of Electronics, Peking University, Beijing 100871, China

[2] The State Key Laboratory of Advanced Optical Communication System and Networks, Department of Electronics, Peking University, Beijing 100871, China

[3] Frontiers Science Center for Nano-optoelectronics, Peking University, Beijing 100871, China

[4] Peng Cheng Laboratory, Shenzhen 518055, China

*Corresponding author. Email:
fzhang@pku.edu.cn (F.Z.); shengwang@pku.edu.cn (S.W.)

**Abstract:** The future interconnect links in intra- and inter-chip require the photodetector with high bandwidth, ultra-wide waveband, compact footprint, low-cost, and compatible integration process with silicon complementary metal-oxide-semiconductor (CMOS) technology. Here, we demonstrate a CMOS-compatible carbon nanotube (CNT) photodetector that exhibits high responsivity, high bandwidth and broad spectral operation over all optical telecommunication band based on high-purity CNT arrays. The ultrafast CNT photodetector demonstrates the 100 Gbit/s Nyquist-shaped on-off-keying (OOK) signal transmission, which can address the demand for high-speed optical interconnects in and between data centers. Furthermore, the photodetector exhibits a bandwidth over 60 GHz by scaling down the active area to 20 μm$^2$. As the CNT photodetectors are fabricated by doping-free process, it also provides a cost-effective solution to integrate CNT photonic devices with CNT-based CMOS integrated circuits. Our work paves a way for future CNT-based high-speed optical interconnects and optoelectronic integrated circuits (OEICs).

Optical interconnect, which has high data transmission speed and low power consumption compared to electrical wires, has developed rapidly in recent years with the tremendous growth of data traffic in communication networks[1-3] and it is the ideal solution for the future OEICs[4-6]. The ultrafast photodetector is an essential component in optical interconnect systems and the bandwidth of photodetector is the main limiting factor for the transmission rate of the communication network[3]. Silicon photonics has been regarded as one of the most promising platforms owing to its mature and low-cost CMOS process and the monolithic OEICs have been demonstrated[4,5]. A silicon-plasmonic detector based on internal photoemission effect (IPE) for 40 Gbit/s data reception has shown a responsivity of 0.12 A/W at 1550 nm and a bandwidth over 40 GHz[7]. However, silicon photodetector is not the ideal candidate for on-chip or off-chip optical interconnect systems due to its indirect bandgap and limited wavelength response range. Waveguide-integrated germanium photodetectors have matured with high responsivity and large bandwidth, but they suffered from a complicated fabrication process with high thermal budget[8,9]. III-V compound-based photodetectors also exhibit high performance, but they are incompatible with the CMOS process owing to lattice mismatch between different materials[10-12].

Over the past few years, two-dimensional (2D) materials based photodetectors including graphene[13], black phosphorus (BP)[14], and transition metal dichalcogenides (TMDs)[15] have greatly aroused the interest of scientists owing to their remarkable electrical and optical properties[16-20]. Graphene-based photodetectors own characteristics of broad spectral response and high operating speed, but they often suffer from low responsivity and large dark current due to the ultra-thin and gapless nature[20]. BP-based photodetectors usually have environmental instability[21] and the largest bandwidth is only about 3 GHz[18]. TMD materials have great promise for high-performance photodetectors due to their strong light-matter interactions. However, they hardly operate at telecom band and usually have low operating speed as a result of large bandgap and low carrier mobility[15].

Carbon nanotubes (CNTs) have shown remarkable electrical and optical properties as a typical one-dimensional material[22], including high carrier mobility ($\sim 10^5$ cm$^2$ V s$^{-1}$) and saturation velocity ($\sim 10^7$ cm s$^{-1}$)[23], ultrasmall intrinsic capacitance, low surface state density, atomic thickness, high absorption coefficient at near infrared band ($\sim 10^5$ cm$^{-1}$)[24], picosecond intrinsic photoresponse time[25] and doping-free CMOS compatible process[26]. Carbon nanotube computer based on complexed CNT circuits has been demonstrated recently[27-29]. Moreover, CNT has great advantage in fabricating monolithic OEICs[30, 31] consisting of light emitter, modulator, silicon waveguide, logic circuit, and photodetector from a single material. Over the past few years, wafer-scaled fabrication of solution-processed high-purity semiconducting CNT films has been developed, which has demonstrated that this process is very efficient[32-34], paving the way to fabricate monolithic OEICs with simple process, low thermal budget and low cost.

In photonics, CNT photodetectors have made great progress by integrated with waveguides[31] and microcavity[35], showing increasing responsivity and detectivity. However, ultrafast photodetectors based on CNT for high-speed optical communications have not been demonstrated to date. The response speed of the reported photodetector based on CNT network film is limited at ~kHz level[30,36]. This is mainly due to the large resistor-capacitor (RC) time delay caused by the large series resistance in the tube-tube junctions and large parasitic capacitance of photodetector in the low resistance Si substrate. Aligned CNT photodetector based on photothermoelectric (PTE) effect have also been reported, however, they also exhibits low response speed (~ kHz level) and low responsivity[37,38]. The realization of ultrafast CNT photodetector for high-speed optical communications still faces challenges.

Here, we report an ultrafast CNT photodetector with a high responsivity ~ 1.5 A/W at 1550 nm and compact active area of 100 $\mu m^2$ based on the well-aligned CNT arrays with high semiconducting purity (>99.9999%)[34]. This ultrafast CNT photodetector shows clear open eye diagrams with 40 Gbit/s non-return-to-zero (NRZ) OOK signals and the bit error rate (BER) is about $5.2 \times 10^{-3}$ under 100 Gbit/s Nyquist-shaped OOK data transmission with digital signal processing (DSP). Moreover, we demonstrate a photodetector that has the highest bandwidth over 60 GHz by scaling down the active area to 20 $\mu m^2$ even at zero bias, paving another route for future high-speed detection with low power consumption and compact footprint.

**RESULT AND DISCUSSION**

The structure of the CNT photodetector is schematically illustrated in Fig. 1a and the detailed fabrication process is described in Methods and Supplementary Fig. S1. The channel material is aligned semiconducting CNT arrays with the density of about 100 ~ 120 tubes/μm and the diameter distribution[34] is about 1.45 ± 0.23 nm. A scanning electron microscope (SEM) image shows clear morphology of CNT aligned arrays (Fig. 1b). The substrate is highly resistive silicon with a SiO$_2$ film of 300 nm to reduce parasitic capacitance. Titanium (Ti)/palladium (Pd) (0.3/80 nm) film and hafnium (80 nm) film is used as the contact electrodes[39,40], respectively, to form asymmetric interdigitated structure to enhance output current. The active area is 100 μm$^2$ with a channel length of 200 nm and a total width of 500 μm. The output signal is collected by the Ti/Au (10/300 nm) electrodes using standard ground-signal-ground (GSG) configuration. A false-color SEM image of the photodetector is shown in Fig. 1c and the Raman spectrum of the CNT arrays is characterized in Fig. 1d. The high G/D ratio of ~ 48 indicating that the defection density of the semiconducting CNT arrays used in this work is very low.

The current-voltage (*I-V*) curves of the photodetector in dark and under illumination with a wavelength of 1550 nm is shown in Fig. 2a. An obvious photocurrent will be generated when it is illuminated by a 1550 nm laser. Fig. 2b shows the output characteristic of the photodetector under different illumination intensities. The photocurrent and photovoltage are, respectively, 0.76 μA and ~ 0.01 V when the illumination power density is 21.3 W/cm$^2$. The short circuit photocurrent follows a linear relationship as a function of the incident power densities (Fig. 2c), indicating that photovoltaic (PV) effect dominates the photoresponse under zero bias. The responsivity is about 34 mA/W at 1550 nm at zero bias. PTE effect would not be dominant in this case, as PTE current is usually generated by an optically induced temperature gradient in suspended CNT or with CNT thick film. And the response speed of devices based on PTE effect is usually slow[37,38]. However, the response speed of our photodetector can reach up to 60 GHz, indicating that PET effect is not dominant.

To investigate the energy band bending in the CNT photodetector at zero bias, a photocurrent mapping experiment was carried out using a 785 nm laser with a spot size of ~ 2 μm. Fig S2a shows the schematic of a CNT photodetector with Pd-Hf asymmetric electrodes. Both the channel length and width of the device are 10 μm. The result in Fig S2b indicates that the obvious photocurrent is mainly generated near the n-type contacted Hf electrode, where a sharp energy band bending (Fig. S2c) assists to dissociate the excitons effectively, same as the previous report[41]. Meanwhile a very small photocurrent can be observed near the p-type contacted Pd electrode because the CNTs are usually p-type due to the adsorption of oxygen or doping by the polymer wrapped on the surface of CNTs.

We find that the responsivity changes rapidly with increasing reverse bias (Fig. 2d) and a large responsivity (~ 1.5 A/W) can be obtained at -1.5 V. It is very different from the traditional photodiode which has a little increasing responsivity at reverse bias comparing with that (34 mA/W) at the zero bias. The possible reason for this rapidly increasing responsivity is photoconductive (PC) effect dominants the photoresponse at reverse bias. The energy band diagram shown in Fig 2e. Under reverse bias, the holes could more easily tunnel from the Hf electrode into the valence band of CNT owing to the thinner Schottky barrier between the Hf electrode and the CNTs. This injection of holes causes an increasing dark current of 11.4 μA at a reverse bias of 0.5 V (Fig. 2a) and the tunneling hole current dominate the dark current. For the average diameter about 1.45 nm in the CNT arrays, the bandgap is estimated ~ 0.5 eV[22]. At larger reverse bias, both electrons and holes can be injected effectively into the channel via tunneling (as shown in Fig. 2e).

The photocurrent gain[42,43] in photoconductor is usually determined by $G = \tau_{lifetime} / \tau_{transit}$, where $\tau_{lifetime}$ is the carrier recombination lifetime and $\tau_{transit}$ is the faster carrier transit time. Here, the photocurrent gain is mainly caused by the amount of holes reinjection at reverse bias and the high mobility of holes with p-type doping in CNT channel (Fig. 2e). The $\tau_{transit}$ for the faster carrier (hole) is estimated about 0.24 ps by $\tau_{transit} = L^2 / \mu V$, where $L \approx 200$ nm is the channel length, $\mu \approx 1120$ cm$^2$ V$^{-1}$ s$^{-1}$ is the mean hole mobility in CNT arrays[44]. $V = 1.5$ V is the bias. The typical carriers recombination lifetime $\tau_{lifetime}$ in the CNTs is larger than 10 ps[45], corresponding normal gain ~ 42, which is close to the experimental result ~ 44. Therefore, the high photocurrent gain is mainly due to the short $\tau_{transit}$ and the long carriers (e-h) recombination lifetime $\tau_{lifetime}$, which results in the bias-dependent responsivity. Therefore, we characterize the electrical and optical properties of the photodetector at a low temperature (Supplementary Fig. S4). The decreasing dark current and responsivity at the low temperature (T = 249K) are owing to less holes injection with low thermal excitation energy and thus a reduction of the photocurrent gain.

The wavelength dependence of the responsivity in the range of 1.2 – 1.8 μm, shown in Fig. 2f, was measured by changing the emission wavelength of a tunable laser. The photoresponse of the CNT photodetector can cover all-optical telecommunication windows from the O band to the U band. The responsivity increases with the increasing excitation wavelength and the maximum is about 85 mA/W (zero bias) at a wavelength of 1.8 μm. The low responsivity of CNT photodetector at zero bias is limited by the relative low absorption of single layer CNT arrays film (~ 4% at 1.8 μm). In principle, the responsivity at zero bias can be further improved with much higher density CNT arrays[35] (~ 250 tubes/μm), multilayer CNT arrays, or integrated with the Si waveguide[31].

In order to characterize the bandwidth performance of the photodetector, we performed the small-signal measurements of optic-electro response ($S_{21}$ parameter) using a 67 GHz vector network analyzer (VNA) with a 50 Ω matched resistance ($R_L$). The measurement setup is shown in Supplementary Fig. S4a (see Methods for a detailed description). As presented in Fig.3a, a 3-dB bandwidth of 22.1 GHz is obtained at a bias of -1.5 V. The optic-electro response is closely related the reverse bias voltages (Fig.3b). With the reverse bias voltage varying from 0 V to 1.5 V, the 3-dB bandwidth of the CNT photodetector gradually increases from 14.5 GHz to 22.1 GHz. The increasing bandwidth is mainly due to the reduced capacitance at a larger reverse bias, leading to the reduction of RC time constant. Therefore, we measured the $S_{11}$ parameter and the representative Smith chart is shown in Supplementary Fig. S5a. The total capacitance ($C$) and resistance ($R_d$) of the CNT photodetector at high frequency were extracted by fitting the real/imaginary part of $S_{11}$ parameters (See Methods for details) and was plotted in Supplementary Fig. S5b. The extracted $C$ is 75 fF and $R_d$ is 78 Ω at zero bias. The parasitic capacitance ($C_p$) is obtained by fitting the $S_{11}$ parameter of a device under an open structure. The measured $C$, $C_p$, and $R_d$ as a function of the voltage bias at high frequency are plotted in Fig. 3c and Supplementary Fig. S5c. Here, $C$ ranges from 70 fF to 143 fF when the voltage bias sweeps from -1 V to 1 V, showing a reduced capacitance with increasing reverse bias. We also find that $C_p$ is a constant of about 62 fF and dominates the total capacitance of photodetector at the reverse bias. An equivalent $RC$ circuit model at high frequency is shown in Fig. 3d. In this model, the total capacitance ($C$) contains CNT junction capacitance ($C_j$) and parasitic capacitance ($C_p$). The total resistance ($R_d$) consists of CNT junction resistance ($R_j$) and series resistance ($R_s$). Generally, the 3-dB bandwidth of a photodetector is mainly limited by transit limit and RC limit. If we assumed the saturation velocity[44] of the carriers in CNT is $3\times10^7$ cm/s, it would take ~ 0.67 ps for the carriers to cross the 200 nm channel length. The transit time-limited bandwidth of the photodetector is about 835 GHz by the equation[20] $f_t = 3.5/2\pi t_{tr}$, where $t_{tr}$ is the transit time. According to the $RC$ model, we calculate the extrinsic $RC$ limited 3-dB bandwidth $1/(2\pi (R_d + R_L) C)$ and intrinsic $RC$ limited 3-dB bandwidth $1/(2\pi R_L C_j)$ (See Methods for details). The calculated extrinsic $RC$ limited 3-dB bandwidth at zero bias is about 16.5 GHz which is consistent with the experimental results of 14.5 GHz (Fig. 3b). The calculated intrinsic $RC$ limited 3-dB bandwidth is larger than 400 GHz at -1 V (Supplementary Fig. S5d), indicating that the bandwidth of our CNT-based photodetectors is mainly limited by the series resistance ($R_d$ ~ 78 Ω at Fig. S5b) and parasitic capacitance (~ 62 fF) at high frequency.

We also evaluated the performance of the 100 μm² CNT photodetector for high-speed optical transmission and the measurement setup is depicted in supplementary Fig. S4b (see Methods for a detailed description). Fig. 4a shows the detected error-free eye diagram for 20 Gbit/s NRZ-OOK signals at different voltages. The signal-to-noise ratio (SNR) is obviously improved by applying a reverse bias, which is mainly due to the increased responsivity and bandwidth at reverse bias. With a bias of -1.5 V, eye diagrams at 30 Gbit/s, 40 Gbit/s, and 60 Gbit/s are also obtained (Fig. 4b). The clear separation between the discrete data levels of the eye diagram at the sampling time indicates low BERs. For NRZ-OOK transmission exceeding 60 Gbit/s, the BER performance is limited by the 59 GHz bandwidth of the analog-to-digital converter (ADC) in our digital storage oscilloscope (DSO). The spectrum components exceeding 60 GHz cannot be captured due to the brick wall filtering characteristic of the DSO. To achieve higher data rates, we adopt Nyquist-shaped OOK signals for transmission (See Methods for details). As shown in Fig. 4c, the BER of 100 Gbit/s Nyquist-shaped OOK signal is about $5.2\times10^{-3}$, which is much lower than the 20% soft-decision forward error correction (SD-FEC) limit[46] of $2.4\times10^{-2}$. This low BER is comparable to the results previously reported by graphene photodetectors[47, 48]. When the data rate is lower than 100 Gbit/s, the mean BERs can be below the 7% hard-decision forward error correction (HD-FEC) limit[49] of $3.8\times10^{-3}$. As the data rates increase to 110 Gbit/s, we can achieve the BER of $2.7\times10^{-2}$. Fig. 4d depicts the measured back-to-back BERs at 60 Gbit/s, 80 Gbit/s, and 100 Gbit/s versus the received optical power. For a signal transmission with a BER of $2.4\times10^{-2}$, the corresponding receiver sensitivities are about -1 dBm, 1 dBm and 5 dBm for 60 Git/s, 80 Gbit/s, and 100 Gbit/s links, respectively. Therefore, our CNT photodetector demonstrates the capability of the detection of high-speed optical signals up to 100 Gbit/s.

As mentioned above, the bandwidth of the photodetector is affected by capacitance, that is to say, we can reduce the capacitance to improve bandwidth of our photodetector. We fabricate total 16 photodetectors with four different device size (200 $\mu m^2$, 100 $\mu m^2$, 40 $\mu m^2$, 20 $\mu m^2$) and Fig. S6 shows the statistical results of capacitance of these devices at zero bias. We can see that the photodetectors with smaller device size have smaller capacitance. For a photodetector with a 20 $\mu m^2$ active area, the total capacitance can be reduced to 33 fF. The representative bandwidth for a photodetector with the active area of 20 $\mu m^2$ at 0 V and -1.5 V is depicted in Fig. 5a, b. It is worth noting that 64 GHz is the measured largest bandwidth for the CNT photodetectors to date, which is sufficient for high-speed optical data transmission with 100 Gbit/s OOK signals as shown in Fig. 4. The photodetector can also operate at zero bias with a large bandwidth of about 60 GHz, indicating that the ultrafast CNT photodetector can be used for high-speed optical communication links with low energy consumption. Fig. S7 shows the *I-V* curves of the 20 $\mu m^2$ photodetector in dark and under illumination with a wavelength of 1550 nm. The bias dependent responsivity and dark current of this photodetector is shown in Fig. 5c. The responsivity can reach up to 1.2 A/W at -0.5 V. Fig. S8 shows the representative dark current characteristic of the ultrafast photodetectors with four different active areas. With the device size scaling from 200 $\mu m^2$ to 20 $\mu m^2$, the dark current decreases significantly. The photodetector with the active area of 20 $\mu m^2$ has the lower dark current below 1$\mu$A at < 0.5 V reverse bias, which allows for the high-speed detection with low energy consumption and compact footprint. The noise equivalent power (NEP) is another significant parameter for characterizing the detection limit of a photodetector. The NEP for the 20 $\mu m^2$ photodetector is estimated about 2 pW/ Hz$^{1/2}$ at zero bias (See Methods for details), indicating a small detectable power for the CNT photodetector.

We also benchmarked the responsivity and bandwidth of our CNT photodetector with state-of-the-art normal incidence (NI) ultrafast photodetectors that have excellent CMOS compatibility (Fig. 5d and Supplementary Tab. S1). It can be seen that our CNT-based photodetector shows the highest bandwidth (> 60 GHz) and remarkable responsivity (1.2 A/W) at the same time. Compared to the conventional Ge photodetectors, our CNT photodetector also has some advantages in broadband operation (1.2 – 1.8 μm) and simple fabrication process with low thermal budget. Meanwhile, the CNT-based photodetector exhibits much lower dark current comparable to graphene-based photodetector. It is worth noting that the high semiconducting purity, well-aligned CNT films can be prepared on 4-inch silicon or quartz wafers with good uniformity[34,44], which meet the requirements for large-scale fabrication CNT photodetectors at different substrates with lower costs compared with the traditional ultrafast photodetectors.

Compared to the waveguide-integrated Ge/III-V/Graphene photodetectors that already have the high bandwidth over 110GHz[9,10,17,50], the bandwidth of CNT NI photodetector is relatively low but it also can be improved with waveguide integration. In the future, the further performance improvement of CNT photodetector requires the optimization of both the devices structure and CNT materials. First, the dark current of CNT photodetector can be further reduced by optimizing the n-type contact quality. Second, the responsivity at zero bias can also can be improved by integrating our CNT photodetector with the silicon waveguide or increasing the density of CNTs. Furthermore, the bandwidth of photodetector and the speed of signal transmission can be improved by reducing the parasitic capacitance and reducing the series resistance by removing the polymer residues wrapped around the CNTs. According to the intrinsic 3-dB bandwidth estimating, the CNT photodetector has the potential to work at 400 GHz, which is a promising candidate for a variety of applications such as high-speed optical communications and microwave photonics.

**CONCLUSION**

In summary, we have reported an ultrafast CNT photodetector with high responsivity ~ 1.5 A/W and compact active area of 100 μm$^2$. The clear open eye diagrams are shown with 40 Gbit/s OOK modulated signal and we achieved data reception at 100 Gbit/s with BER of about $5.2 \times 10^{-3}$. A largest bandwidth of 64 GHz has been achieved by reducing the active area of the device to 20 μm$^2$. Moreover, it shows a relatively high responsivity of 90 mA/W and a high bandwidth ~ 60 GHz even at zero bias. This CMOS-compatible CNT ultrafast photodetector with advantages of high responsivity, broad wavelength range, high bandwidth, high-speed data reception, compact footprint, and simple fabrication process, has the great potential for future application in high-speed optical communication systems.

**METHODS**

**Fabrication of the ultrafast CNT photodetector.**

The aligned CNT arrays used in this work are purchased from Beijing Huatan Yuanxin Electronic Technology Company. The 4' wafer CNT film are fabricated by the multiple dispersion and dimension-limited self-alignment (DLSA) procedure[34]. The aligned CNT arrays are treated with yttrium oxide coating and decoating (YOCD) processes[51] before device fabrication. Then, CNT film is transferred to high-resistivity $SiO_2$/Si substrates by wet transfer process with HF solution (volume ratio of 1:6) to reduce parasitic capacitance. In the next fabrication process, firstly, the channel area is defined by electron beam lithography (EBL, Raith 150), followed by inductively coupled plasma etching (ICP) to remove the extra CNTs. Secondly, Ti/Pd bilayer films of 0.3/80 nm are deposited on the CNT array films by electron beam evaporation (EBE, DE400) to form a p-type contact electrode. Thirdly, Hf film of 80 nm is deposited on the CNT array films by a sputtering method (PVD 750, Kurt J. Lesker) to form the n-type electrode. Finally, Ti/Au bilayer films of 10/300 nm are deposited as GSG pads by EBE. In each EBL process, polymethyl methacrylate (PMMA) is used as positive resist and all the metal films were deposited with a standard lift-off process.

**Electrical and optoelectronic characterization of the ultrafast CNT photodetector.**
A 488 nm continuous wavelength laser is used for Raman spectroscopy characterization (Jobin Yvon/Horiba Company). The electronic transportation characteristic is measured using a commercial semiconductor analyzer (Keithley 4200). Optoelectronic measurements are carried out using a super-continuous spectrum laser (NKT Company), which has an IR wavelength range from 1165 nm to 2100 nm and has a power density of 21.3 $W/cm^2$ at 1550 nm. The mapping photocurrent profile was performed using a 785 nm laser. The characteristic of the CNT photodetector at low temperature were measured in the vacuum chamber which can be cooled by liquid nitrogen. The devices are encapsulation with 180 nm PMMA during the electrical and optoelectronic test to improve the stability.

**High-frequency measurements of the ultrafast CNT Photodetector.**

The bandwidth of the photodetectors are characterized using a commercial 67 GHz vector network analyzer (Agilent 8703) and a lightwave component analyzers (LCA). A continuous wave laser at 1550 nm and a $LiNbO_3$ intensity modulator are built in the LCA. The modulator is typically tested for their frequency response in advance, which is built in the system. When the photodetector under test is connected, the VNA will automatically subtract the response of the modulator from the previous test to obtain the real response of our photodetector. The modulated optical output from the LCA is adjusted by a polarization controller (PC) and then coupled into the CNT photodetector through a standard single-mode fiber. The detected signal is extracted by a ground-signal-ground (GSG) microwave probe, amplified by an electrical amplifier (SHF S804B) and subsequently fed back to the VNA. A bias-tee (SHF BT65 B) is used to provide a voltage bias to the photodetector.

**The RC calculation of the ultrafast CNT photodetector.**

The reflection coefficient $S_{11}$ was measured using a 3 mm S-parameter measurement system with a VNA (N5245A) up to 50 GHz. The high-frequency impedance $Z_{11}$ of the photodetector can be calculated according to $Z_{11} = Z_0 ((1+S_{11})/ (1-S_{11}))$, where $Z_0$ is the 50 Ω system impedance. Based on the equivalent circuit model, the total high-frequency impedance $Z_{11}$ can be described by $Z_{11} = R_d + |X_c| i$, where $R_d = R_j + R_s$ is the device resistance at high frequency, $|X_c| = 1/2\pi f (C_p+C_j)$ is the capacitive reactance of the device at high frequency. Therefore, $R_d$, $C_p$, and $C_j$ can be obtained from $Z_{11}$ based on the device under testing (DUT) and open structure. The extrinsic 3-dB bandwidth can be calculated using $f_{3dB\text{-}extrinsic} = 1/(2\pi (R_d + R_L)(C_p + C_j))$. Similarly, the intrinsic 3-dB bandwidth can be calculated to evaluate the potential of our photodetector using $f_{3dB\text{-}intrinsic} = 1/(2\pi R_L C_j)$ when we neglect the $R_d$ and $C_p$ under ideal conditions.

**High-speed data detection of the ultrafast CNT photodetector.**

An external cavity laser (ECL) at 1550nm is utilized as the optical source. The arbitrary waveform generator (AWG) (Keysight M8194A) operating at 120GSa/s generates 20/30/40/60 Gbit/s NRZ-OOK signals for eye diagrams measurements and 80/90/100/110 Gbit/s Nyquist OOK signals for BER measurements. The Nyquist shaping enables an approximately rectangle signal spectrum equal to half of its baud rate. The high-speed signals are modulated by a commercial LiNbO$_3$ intensity modulator (with a bandwidth 30 GHz) biased at quadrature point. An erbium-doped optical fiber amplifier (EDFA) is used to amplify the modulated optical signal. A variable optical attenuator (VOA) was used to attenuate optical power to measure the receiver sensitivity of the photodetector in the BER measurements. After adjusting the polarization state by a polarization controller (PC), the modulated optical signal with an estimated power of 15 dBm is fed into the CNT photodetector. Considering the interdigital device structure and spot size of the single-mode fiber, the actual light power irradiated on active area of the photodetector is about 7 dBm according to $P_{CNT}$ $= P_0 * S / \pi r^2$, where $P_0 = 15$ dbm is the power of the modulated optical signal, $S = 100$ μm$^2$ is the active area of the CNT photodetector, $r \approx 15$ μm is the spot size of the single-mode fiber. A bias-T (SHF BT65 B) is used to apply a DC bias to the photodetector. Then, the detected signals are amplified by an electrical amplifier (SHF S804 B) and subsequently sampled by a digital storage oscilloscope (Keysight UXR0594AP) with a 59 GHz bandwidth. A series of digital signal processing steps, including root raise cosine (RRC) filter, synchronization and train sequence-based time domain equalization, were employed in the off-line processing of the signals to optimized the BER performance. The inter-symbol-interference due to the bandwidth limitation of the commercial modulator and the CNT photodetector is mitigated by channel equalization based on the DSP algorithms.

**The NEP calculation of the ultrafast CNT photodetector.**

NEP can be calculated by the equation $NEP = P_{in}/(I_{ph}/I_n)$, where $P_{in}$ is the laser power that reaches the device channel, $I_{ph}$ is the output photocurrent at zero bias, $I_n$ is the niose current of the photodetector. Because our photodetector works at zero bias, the noise mainly originates form Johnson niose $I_n = (4kT \triangle f /R)^{1/2}$, where $k$ is Boltzmann's constant, $T$ is the temperature and $R$ is the differential resistance of the PD at zero bias. According to this, we calculated that $R$ is 0.65 MΩ，NEP is 2 pW/ Hz$^{1/2}$.

**ASSOCIATED CONTENT**

**Additional Information**

Supplementary Materials, Figs. S1 to S8，Tables S1.

**AUTHOR INFORMATION**

**Corresponding Author**

fzhang@pku.edu.cn (F.Z.)
shengwang@pku.edu.cn (S.W.)

**Acknowledgements**

This work was supported by the National Key Research & Development Program (Grant No. 2020YFA0714703), National Science Foundation of China (Grant No. 62071008, U21A6004) and the major key project of Peng Cheng Laboratory (PCL 2021A14).

**Author contributions**

S.W. proposed and supervised the project. S. W. designed the experiment. W.W. performed the fabrication and measurement of the CNT photodetectors. F.Y. was helpful for the bandwidth measurements, X.F. was responsible for eye diagrams measurements and transmission experiments. X.C. and X.L. are helpful for the Hf film deposition. F.Z. guided the high-speed optical transmission experiments for validating the CNT photodetectors. W.W. and S.W. analyzed the data and co-wrote the manuscript. F.Y., X.F. and F. Z. contribute to the data analysis and wrote the transmission experiments part of the manuscript. All authors discussed the results and contributed to manuscript preparation.

**Competing Interests**

The authors declare no competing interests.

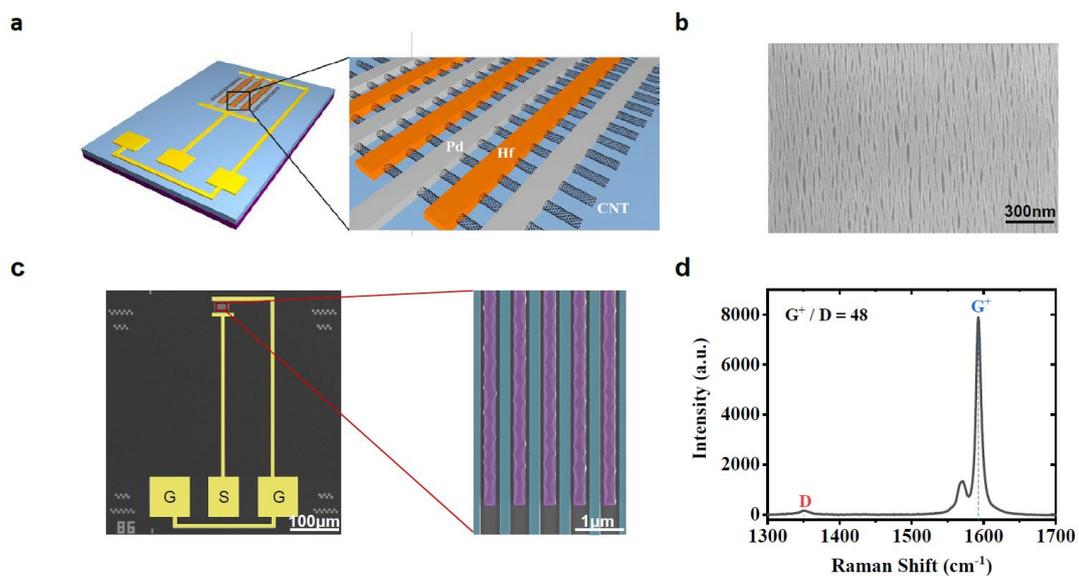

**Figure 1 | Structure and characterization of the ultrafast CNT Photodetector. a,** Schematic of the interdigitated asymmetrically contacted CNT photodetector. **b,** SEM characterization of aligned semiconducting CNT arrays. **c,** False-color SEM image of the photodetector with GSG pads. **d,** Raman spectrum characterization of aligned CNT arrays with an excitation wavelength of 488 nm with the polarization parallel to the CNT arrays.

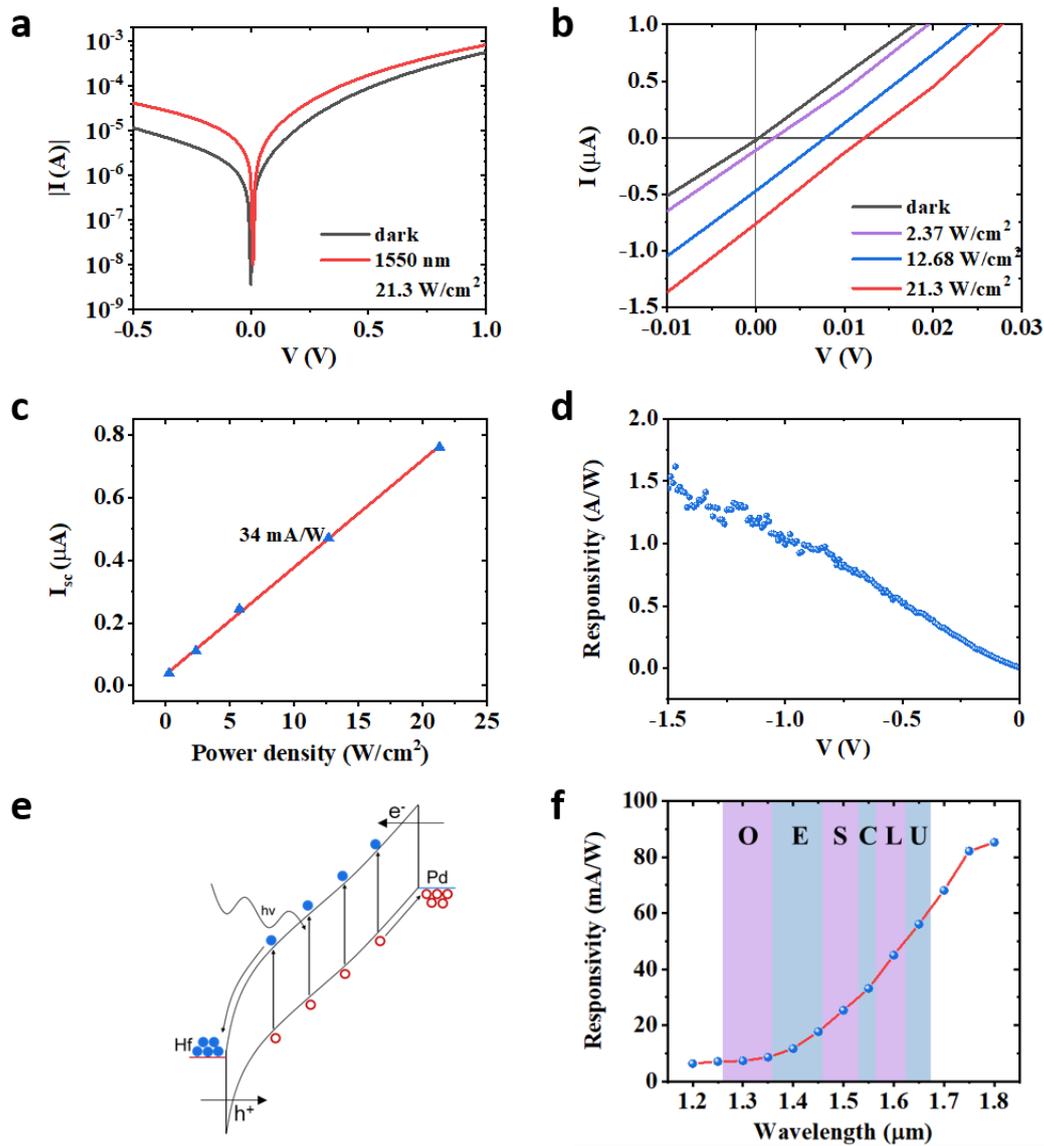

**Figure 2 | Electrical and optical characteristics of the ultrafast CNT Photodetector. a,** Current versus bias in dark and under illumination at an excitation wavelength of 1550 nm with the polarization parallel to the CNT arrays. The illumination power density is 21.3W/cm$^2$. The active area of photodetector is 100 μm$^2$. **b,** Output curves under different incident power intensities. **c,** Experimental (blue triangle) and fitting (red line) results of photocurrent (V = 0 V) versus the incident power densities at λ = 1550 nm. **d,** Reverse bias dependent responsivity of the photodetector. **e,** Band diagram of the device under large reverse bias. **f,** Wavelength dependence of the responsivity of CNT photodetector from 1.2 μm to 1.8 μm at V = 0 V. O band (1260-1360nm). E band (1360-1460nm). S band (1460-1530nm). C band (1530-1565nm). L band (1565-1625nm). U band (1625-1675nm).

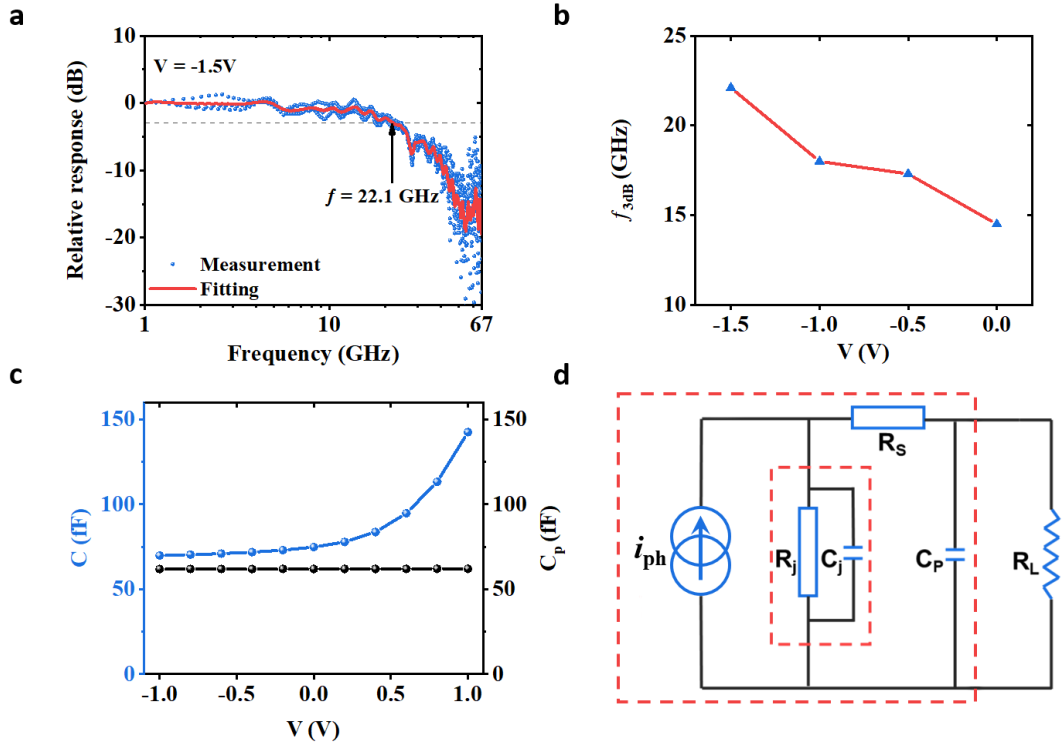

**Figure 3 | Dynamic characteristics of the ultrafast CNT Photodetector. a,** Experimental (blue dots) and fitting (red curve) frequency response at -1.5 V. The active area of photodetector is 100 μm². **b,** The measured 3-dB bandwidth at different voltage biases. **c,** Bias dependence of the total capacitance ($C$) and parasitic capacitance ($C_p$). **d,** Equivalent $RC$ circuit model of the CNT photodetector at high frequency. $R_j$, $C_j$, $R_s$, $C_p$, and $R_L$ represent the CNT junction resistance, junction capacitance, series resistance, parasitic capacitance, and 50 Ω load.

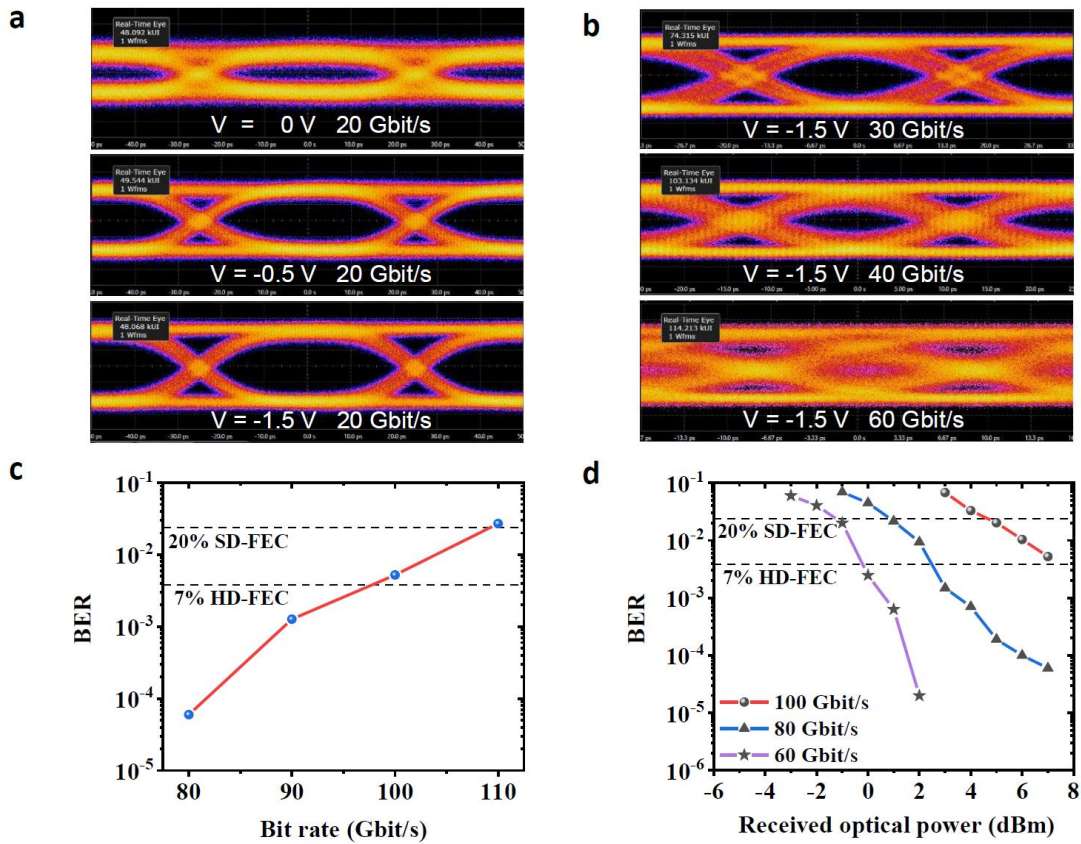

**Figure 4 | High-speed data reception of the ultrafast CNT Photodetector with the active area 100 μm².** **a,** Received eye diagrams at the data rate of 20 Gbit/s when the photodetector was biased at 0 V, -0.5 V, and -1.5 V. The active area of photodetector is 100 μm². **b,** Eye diagrams for 30 Gbit/s, 40 Gbit/s, and 60 Gbit/s links at the bias of -1.5 V. **c,** Measured BER versus bit rates for NRZ-OOK-signals from 80 Gbit/s to 110 Gbit/s. **d,** Measured of BER versus the received optical power for 60 Gbit/s, 80 Gbit/s, and 100 Gbit/s OOK signals. The corresponding sensitivities are about -1 dBm, 1 dBm, and 5 dBm, respectively.

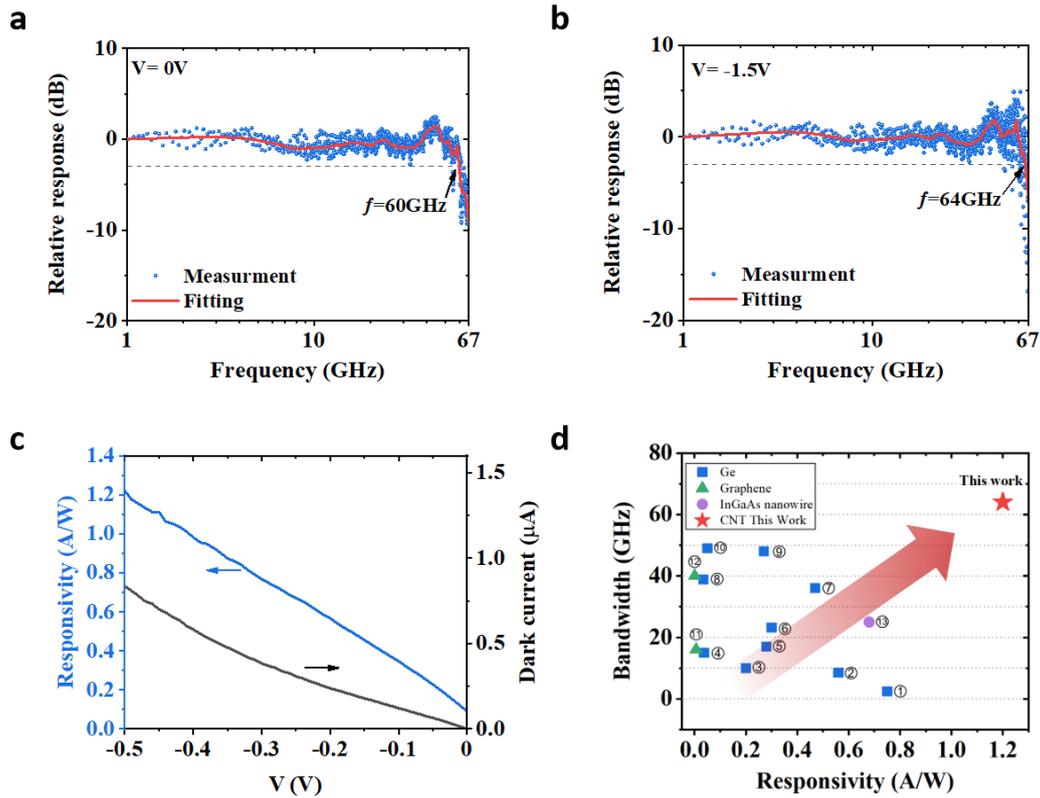

**Figure 5 | The performance of the ultrafast CNT photodetector with active area of 20 μm$^2$. a, b,** Experimental (blue dots) and fitting (red curve) relative response as a function of the modulation frequency at 0 V and -1.5 V for a photodetector with an active area of 20 μm$^2$. **c,** The responsivity and dark current as a function of the reverse bias for the 20 μm$^2$ photodetector. **d,** Benchmarking of the responsivity and bandwidth of our CNT photodetector with state-of-the-art normal incidence ultrafast photodetectors that have excellent CMOS compatibility. All data used here are listed in Supplementary Tab. S1.